\documentclass[aps,prd,preprint,nofootinbib,a4paper,11pt,superscriptaddress]{revtex4-1}

\usepackage[T1]{fontenc}
\usepackage[centertags]{amsmath}
\usepackage{amsfonts}
\usepackage{amssymb}
\usepackage[sort&compress]{natbib}% упорядочивает ссылки
\usepackage[hypertex]{hyperref}%hypertex
\DeclareMathOperator{\Tr}{Tr}
\DeclareMathOperator{\re}{Re}

\DeclareMathOperator{\sgn}{sgn}

\newcommand{\lan}{\langle}
\newcommand{\ran}{\rangle}

\newcommand{\spx}{\mathbf{x}}

\newcommand{\br}[1]{(#1)}
\newcommand{\Zs}{\textsf{Z}}

\newcommand{\e}{\varepsilon}

\newcommand{\s}{\sigma}

\newcommand{\al}{\alpha}
\newcommand{\be}{\beta}
\newcommand{\ga}{\gamma}
\newcommand{\Ga}{\Gamma}
\newcommand{\de}{\delta}

\newcommand{\la}{\lambda}

\begin{document}
\allowdisplaybreaks[4]

\title{High-temperature expansion of the one-loop free energy of a scalar field on a curved background}

\date{\today}

\author{I.S. Kalinichenko}
\email[E-mail:]{theo@sibmail.com}
\affiliation{Physics Faculty, Tomsk State University, Tomsk, 634050 Russia}

\author{P.O. Kazinski}
\email[E-mail:]{kpo@phys.tsu.ru}
\affiliation{Physics Faculty, Tomsk State University, Tomsk, 634050 Russia}
\affiliation{Department of Higher Mathematics and Mathematical Physics, Tomsk Polytechnic University, Tomsk, 634050 Russia}

\begin{abstract}

The complete form of the high-temperature expansion of the one-loop contribution to the free energy of a scalar field on a stationary gravitational background is derived. The explicit expressions for the divergent and finite parts of the high-temperature expansion in a three-dimensional space without boundaries are obtained. These formulas generalize the known one for the stationary spacetime. In particular, we confirm that for a massless conformal scalar field the leading correction to the Planck law proportional to the temperature squared turns out to be nonzero due to non-static nature of the metric. The explicit expression for the so-called energy-time anomaly is found. The interrelation between this anomaly and the conformal (trace) anomaly is established. The natural simplest Lagrangian for the ``Killing vector field'' is given.

\end{abstract}

\pacs{04.62.+v}
%04.62.+v QFT in curved spacetime

\maketitle

\section{Introduction}

The high-temperature expansions of the partition functions are the classical subject of quantum field theory on a curved background. As for the one-loop contributions to the free energy of quantum fields, we just mention the works \cite{DowKen,NakFuk,DowSch,DowSch1,HuCrSt,Kirsten,Campor,GusZeln,Fursaev1,Fursaev2} where the different approaches to this problem were implemented. Surprisingly, in spite of the fact that the first attempts to find the high-temperature expansion were undertaken over about thirty years ago \cite{DowKen}, the problem was not completely solved, to our knowledge, in its general statement for the arbitrary stationary (non-static) gravitational background. The present paper is aimed to fill this gap. Such a high-temperature expansion for the free energy will be derived here. In particular, for a massless conformal scalar field we shall obtain the leading correction to the Planck law proportional to the temperature squared and confirm the result of \cite{Fursaev1,Fursaev2}. This correction is absent for a static gravitational background \cite{DowKen,NakFuk,DowSch}. Also we shall derive the explicit expression for the finite part of the high-temperature expansion, which seems to be a new result.

Apart from the immediate implications for the Casimir effect, astrophysics, and cosmology, the high-temperature expansion of the free energy for fermions can be used to analyze the derivative expansion of the one-loop contribution to the effective action at zero temperature (see, e.g., \cite{olopq,gmse}) regularized by the energy cutoff. According to the general prescriptions of the renormalization theory \cite{Collins}, the structures appearing as the divergencies in the effective action must be included to the initial action to cancel these divergencies (not necessarily to zero). As a rule, these divergencies and the finite part depend nontrivially on the Killing vector $\xi^\mu$ defining the stationarity of the background and the vacuum state of the quantum fields (\cite{psfss}, see also \cite{DeWQFTcspt,BrOtPa,FrZel}). We shall see in the present paper that this is indeed the case. Hence, the analysis of the high-temperature expansion sheds a light on the low energy quantum dynamics of the vector field $\xi^\mu$ as it was discussed in \cite{gmse,psfss}. To see how this works, one may bear in mind the expansion of the Heisenberg-Euler effective action \cite{HeisEul,Schwing}, where the first non-trivial term describes the light by light scattering.

Another interesting point following from the results of the present paper is the interrelation between the conformal (trace) anomaly and the energy-time anomaly (the notion of the latter was introduced in \cite{psfss}, see also below). As we shall see, we cannot renormalize the quantum theory of a massless conformal scalar field in such a way that both the conformal and energy-time anomalies vanish. The elimination of the one anomaly results in the appearance of the other and vice versa. It is noteworthy that the factor at the logarithm of the temperature entering the high-temperature expansion and determining the anomaly turns out to be independent of the Killing vector and coincides with the standard expression for the conformal anomaly \cite{DowSch,Fursaev2}. However, the finite part of the high-temperature expansion does involve certain contractions of the Killing vector field. Loosely speaking, the problem of dependence of the effective action on the Killing vector comes from the infrared modes of the gravitational field and so it is reasonable that the renormalization group beta function defining the ultraviolet behaviour of the theory is independent of the Killing vector. Though, of course, this fact is not a priori obvious since both in the ultraviolet and the infrared regions we have the mode functions defined with respect to the same Killing vector (see, however, \cite{Fursaev2} for the possible proof).

The paper is organized as follows. Since the subject of the article is rather technical, we partition it into the small sections distinguishing the major successive steps of the derivation, which culminates in Sec. \ref{ResImpl} where the results and its implications are discussed. In Sec. \ref{GenForm}, we provide general formulas for the one-loop correction to the free energy. The main technical tool, which we shall employ to derive the high-temperature expansion, is the heat kernel of a Laplacian operator (see \cite{VasilHeatKer} for review). Therefore, in Secs. \ref{Red}, \ref{ResHKE}, and \ref{EvaInt} we reduce our problem to the evaluation of the heat kernel expansion coefficients. At first, Sec. \ref{Red}, we reduce the three-dimensional problem to the four-dimensional one to provide the explicit general covariance to the expansion. The method, which is used here, is a finite dimensional analog of the gauge fixing procedure in the functional integral. Then, in Sec. \ref{ResHKE}, we resum the heat kernel expansion applying the theorem proved in \cite{Parker}. Notice that we do not use the conformal transformation to derive the high-temperature expansion \cite{DowKen,DowSch,DowSch1,Kirsten,Campor,GusZeln,Fursaev1,Fursaev2}, but apply the method directly to the Fourier transformed Klein-Gordon operator. So, our approach is rather close to the one used in \cite{NakFuk}. After that, the problem becomes in essence the same as in a flat spacetime. In Sec. \ref{EvaInt}, we evaluate the expansions of necessary integrals employing the procedure used in \cite{olopq}. Then, in Sec. \ref{HTE}, a general formula for the high-temperature expansion of the free energy is obtained. In Sec. \ref{ResImpl}, we particularize the general formulas to the three-dimensional space and single out the divergent and finite parts of the high-temperature expansion. At this point we essentially employ the results of \cite{Ven}, where the heat kernel coefficients were derived up to $a_6$. In Appendix \ref{4_Kill}, we give some useful relations for the metric possessing the Killing vector. In Appendix \ref{HKE_coeff}, the relevant parts of the heat kernel coefficients borrowed from \cite{VasilHeatKer,Ven,Avramid,Gilkey} are calculated.

Despite the fact that we study the high-temperature expansion of the free energy of quantum scalar field on a gravitational background, the results are easily generalized to higher spins and other stationary backgrounds. The problem lies only in the amount of calculation, which increases due to the additional terms in the heat kernel expansion. A non-zero chemical potential can be also included. We postpone the investigation of these problems for the future research.

We shall use the following conventions for the curvatures and other structures appearing in the heat kernel expansion:
\begin{equation}
    R^\al_{\ \be\mu\nu}=\partial_{[\mu}\Ga^\al_{\nu]\beta}+\Ga^\al_{[\mu\ga}\Ga^\ga_{\nu]\beta},\qquad R_{\mu\nu}=R^\al_{\ \mu\al\nu},\qquad R=R^\mu_\mu.
\end{equation}
The square and round brackets at a pair of indices denote antisymmetrization and symmetrization without $1/2$, respectively. The Greek indices are raised and lowered by the metric $g_{\mu\nu}$ which has the signature $-2$. Also we assume that the metric possesses the timelike Killing vector $\xi^\mu$:
\begin{equation}
    \mathcal{L}_\xi g_{\mu\nu}=0,\qquad\xi^2=g_{\mu\nu}\xi^\mu\xi^\nu>0.
\end{equation}
The space dimension will be denoted by $d$ and $d=3$. Nevertheless, we shall put $d=3$ only in the final result. The system of units is chosen such that $c=\hbar=1$.

\section{General formulas}\label{GenForm}

Consider a scalar quantum field on a stationary gravitational background at a finite reciprocal temperature $\beta$. The free energy for this system is defined in the standard way
\begin{equation}\label{free_energy}
    e^{-\beta F}:=\Tr e^{-\be \mathcal{H}},
\end{equation}
where $\mathcal{H}$ is the Hamiltonian of the scalar field expressed in terms of the creation-annihilation operators. The mode functions of this field corresponding to the energy $\omega$ span the kernel of the Klein-Gordon operator,
\begin{multline}\label{KG_eq}
    H(x,y)=(-\nabla^2_x-m^2)\frac{\de(x-y)}{|g|^{1/4}(x)|g|^{1/4}(y)}=\\
    =|g|^{-1/4}(x)\biggl[-|g|^{-1/4}(x)\partial_\mu\sqrt{|g|}g^{\mu\nu}\partial_\nu|g|^{-1/4}(x)-m^2\biggr]\frac{\de(x-y)}{|g|^{1/4}(y)},
\end{multline}
where all the time derivatives should be replaced by $-i\omega$. This operator, which we denote as $H(\omega)$, must be supplemented by the appropriate boundary conditions. To simplify further calculations we assume that the system considered is large enough to neglect the boundary effects or the space represents a compact manifold without boundary. The operator $H(\omega)$ is Hermitian with respect to the measure $\sqrt{|g|}$ on the square-integrable functions depending on $\spx$.

Our aim is to calculate the one-loop contribution to the free energy \eqref{free_energy}. To this end, we use the fact that
\begin{equation}\label{spectr_dens}
    \partial_\omega\Tr_d\theta(H(\omega))=\sum_k\sgn(\e'_k(\omega))\de(\omega-\omega_k)
\end{equation}
defines the signed spectral density in the $\omega$-space for the self-adjoint operator $H(\omega)$ possessing the eigenvalues $\e_k(\omega)$. Here the Heaviside step function $\theta(H(\omega)-\e)$ is the spectral decomposition of unity associated with $H(\omega)$ and $\omega_k$ is the solution to the equation
\begin{equation}
    \e_k(\omega)=0.
\end{equation}
If this equation has several solutions then the sum over all such solutions should be taken in \eqref{spectr_dens}. Note that $\omega$ may be any parameter of the self-adjoint operator and not only the energy. In the case when $\omega$ is the energy as defined above, $\e'_k(\omega)$ is usually positive for $\omega>0$ (the particle branch) and negative for $\omega<0$ (the antiparticle branch). Differentiating the equations
\begin{equation}
    H(\omega_k,m^2)\psi_k(m^2)=0,\qquad\lan\psi_k|\psi_k\ran=1,
\end{equation}
with respect to $m^2$, it is easy to see that the last statement is valid for the same cases, when the standard prescription $m^2\rightarrow m^2-i0$ defines the Feynman propagator. It follows from the quasiclassical asymptotic of the operator $H(\omega)$, where all the derivatives entering it are replaced by the momenta $-ip_\mu$, that $\e'_k(\omega)$ can be negative for $\omega>0$ under the ergosphere, i.e., the antiparticle branch appears in the positive-frequency region. This results in the particle creation \cite{Starobin} and the reconstruction of the vacuum state. Therefore we restrict our consideration by the region out of the ergosphere.

Assuming that the lowest particle energy is strictly positive (this is the case when the system is placed in a finite ``box''), we can write the one-loop contribution to the free energy from the particles as
\begin{equation}\label{om_potent}
    \mp\be F=\int_0^\infty d\omega\partial_\omega\Tr_d\theta(H(\omega))\ln(1\pm e^{-\be\omega})=\pm\beta\int_0^\infty d\omega\frac{\Tr_d\theta(H(\omega))}{e^{\be\omega}\pm1},
\end{equation}
where plus corresponds to fermions and minus is for bosons. The contribution from the antiparticles is given by the same integral, but with $H(-\omega)$ instead of $H(\omega)$. Hence, the one-loop contribution to the effective action from one bosonic mode at zero temperature reads (see, e.g., \cite{olopq,gmse})
\begin{equation}\label{one_loop_bos}
    \Ga^{\br1}_{1b}/T=-\lim_{\beta\rightarrow0}\partial_\be(\be F)=\partial_\be\Big[\beta\int_0^\infty d\omega\frac{\Tr_d\theta(H(-\omega))}{e^{\be\omega}+1}\Big]_{\be\rightarrow0},
\end{equation}
where $\be^{-1}$ plays the role of the energy cutoff and $T$ is the time interval tending to infinity. The fermionic free energy for a scalar field (i.e. \eqref{om_potent} with the plus sign) can be also used to estimate the partition function for fermions when the spin-gravity interaction is negligible. For further purposes it is useful to represent the step function in the integral form:
\begin{equation}\label{theta}
    \theta(H(\omega))=\int_{C}\frac{d \tau}{2\pi i} \frac{e^{-\tau H(\omega)}}{\tau},
\end{equation}
where the contour $C$ runs along the imaginary axis from top to bottom and encircles the origin from the left. The exponent on the right-hand side is the so-called heat kernel. In order to obtain the high-temperature expansion ($\be\rightarrow0$), it will be sufficient to know the expansion of the trace of the heat kernel in a series in $\tau$.

\section{Reduction to $d+1$}\label{Red}

Of course, we can apply the formulas from \cite{VasilHeatKer,Avramid,Ven,Gilkey} directly to \eqref{theta}, but this leads to rather cumbersome expressions that are not explicitly covariant because of the $(3+1)$ decomposition. Therefore, we reduce our $3$-dimensional problem to the $4$-dimensional one preserving explicitly the general covariance. To this end, we introduce the auxiliary integration variable,
\begin{equation}\label{deltafunction}
    e^{-\tau H(\omega)}=\int_{-\infty}^{\infty}d p_0 e^{-\tau H(p_0)}\delta(p_0-\omega)=
    \int_{-\infty}^{\infty}d p_0 e^{-\tau [H(p_0)+\frac{1}{\xi^2}(\omega^2-p^2_0)]}\delta(p_0-\omega),
\end{equation}
and employ the standard Gaussian representation of the delta function
\begin{equation}
    \delta(p_0-\omega)=\lim_{\la\rightarrow-\infty}\sqrt{\frac{\la}{\pi}}e^{-\tau\la(p_0-\omega)^2}\tau^{1/2}.
\end{equation}
The second term in the exponent in \eqref{deltafunction} tends to zero when $\la$ goes to minus infinity. The usefulness of such an addition will become clear soon. The expressions with a finite $\la$ correspond to the same system but with a Gaussian broadening of the spectrum. Denoting
\begin{equation}
    H_\la(p_0,\omega):=H(p_0)+\la(p_0-\omega)^2+\frac{1}{\xi^2}(\omega^2-p^2_0),
\end{equation}
we have
\begin{equation}
    \Tr_d e^{-\tau H(\omega)}=\lim_{\la\rightarrow-\infty}\sqrt{\frac{\la}{\pi}}\int_{-\infty}^{\infty}d p_0 \Tr_d e^{-\tau H_{\la}(p_0,\omega)}\tau^{1/2}
    =\frac1{T}\lim_{\la\rightarrow-\infty}2\sqrt{\pi\la}\Tr_D e^{-\tau H_{\la}}\tau^{1/2},
\end{equation}
where $D:=d+1$ is the spacetime dimension. Then we make a similarity transform such that the diagonal of the heat operator in the $x$-representation and, consequently, its the trace are left intact, while the Hamiltonian becomes (see also Appendix \ref{4_Kill} for the notation)
\begin{equation}\label{H_G}
    H_G:=(\xi^2)^{1/4}H_\la(\xi^2)^{-1/4}=-G^{\mu\nu}(\tilde{\nabla}_{\mu}-i\omega g_\mu)(\tilde{\nabla}_{\nu}-i\omega g_\nu)-X,
\end{equation}
where $g_\mu:=\xi_\mu/\xi^2$ and
\begin{equation}\label{metric}
\begin{gathered}
    G_{\mu\nu}=g_{\mu\nu}-\xi^2g_\mu g_\nu+\la^{-1}g_\mu g_\nu,\qquad G^{\mu\nu}=g^{\mu\nu}-\xi^2(1-\la \xi^2)g^\mu g^\nu,\qquad \det G_{\mu\nu}=\la^{-1}g^2\det g_{\mu\nu},\\
    X=V-E,\qquad V=\frac14h_\mu h^\mu-\frac12\nabla^\mu h_\mu,\qquad E=\frac{\omega^2}{\xi^2}-m^2,
\end{gathered}
\end{equation}
and $h_\mu:=\partial_\mu\ln\sqrt{\xi^2}$. The connection $\tilde{\nabla}_\mu$ is the Levi-Civita connection compatible with the metric $G_{\mu\nu}$. The latter is negative definite  so that the operator $H_G$ is of the Laplacian type.

\section{Resummation of the heat kernel expansion}\label{ResHKE}

Now we can use the heat kernel expansion
\begin{equation}\label{HKE_ini}
    \lan x|e^{-\tau H_{G}}|y\ran=\sum_{k=0}^{\infty}a_k(x,y)\frac{\tau^{k-D/2}}{(4\pi)^{D/2}}.
\end{equation}
It is also relevant here that the space has no boundary. Otherwise the terms with the half-integer $k$ appear. To exploit the heat kernel expansion, we resum it using the theorem proved in \cite{Parker}
\begin{equation}\label{heat_kern_res}
    \lan x|e^{-\tau H_{G}}|y\ran=e^{\tau[X(x)+\frac16\tilde{R}(x)]}\lan x|e^{-\tau[X(x)+\frac16\tilde{R}(x)]}e^{-\tau H_{G}}|y\ran
    =e^{\tau[X(x)+\frac16\tilde{R}(x)]}\sum_{k=0}^\infty \tilde{a}_k(x,y)\frac{\tau^{k-D/2}}{(4\pi)^{D/2}},
\end{equation}
where $\tilde{a}_k$ are obtained from $a_k$ by removing the terms containing $X$ and $\tilde{R}$ without derivatives. As for the $X$ terms, this theorem simply follows from the observation that
\begin{equation}
    \lan x|e^{-\tau[X(x)+\frac16\tilde{R}(x)]}e^{-\tau H_{G}}|y\ran
\end{equation}
is invariant under the transform $X\rightarrow X+const$ and so its expansion does not contain $X$ without derivatives. On the other hand, expanding
\begin{equation}
    \lan x|e^{-\tau[X(x)+\frac16\tilde{R}(x)]}e^{-\tau H_{G}}|y\ran=e^{-\tau[X(x)+\frac16\tilde{R}(x)]}\lan x|e^{-\tau H_{G}}|y\ran
\end{equation}
in $\tau$, we see that at the same power of $\tau$:
\begin{equation}
    \tilde{a}_k=a_k+O(X,\tilde{R}),
\end{equation}
where $O(X,R)$ are the terms containing $X$ and $\tilde{R}$ without derivatives and $a_s$ with $s<k$. If $\tilde{a}_k$ does not contain $X$ and $\tilde{R}$ without derivatives in virtue of summation of such terms to the exponent then $\tilde{a}_k$ is obtained by a mere obliteration of the terms in $a_k$ proportional to $X$ or $\tilde{R}$ without derivatives. The fact that $\tilde{a}_k$ do not depend on the scalar curvature $\tilde{R}$ can be proven by a direct inspection of the heat kernel expansion coefficients \cite{Parker} or by solving the defining equations for the heat kernel in the Gaussian approximation (see, e.g., \cite{psfss}).

\section{Evaluation of the integrals over $\tau$ and $\omega$}\label{EvaInt}

For brevity, we introduce the notation
\begin{equation}
\begin{gathered}
    \tilde{m}^2:=m^2+V+\frac16\tilde{R}=m^2+\frac14h^2+\frac16(R-\nabla^\mu h_\mu)-\frac1{24}(\la^{-1}-\xi^2)f^2=\\
    =m^2+\frac14h^2+\frac16[R-(\la^{-1}-\xi^2)R_{\mu\nu}g^\mu g^\nu-\la^{-1}g^2\nabla^\mu h_\mu],
\end{gathered}
\end{equation}
where we have used the relations from Appendix \ref{4_Kill}. The quasiclassical expansion for the free energy \eqref{om_potent} after the above transformations becomes
\begin{equation}\label{om_pot_exp1}
    -F=\frac{-i}{T}\lim_{\la\rightarrow-\infty}\sqrt{\frac{\la}{\pi}}\sum_{k=0}^\infty\int dx\sqrt{G}\int_0^\infty \frac{d\omega \tilde{a}_k(\omega,x)}{e^{\be\omega}\pm1}\int_C\frac{d\tau}{(4\pi)^{D/2}}e^{-\tau(\frac{\omega^2}{\xi^2}-\tilde{m}^2)}\tau^{k-d/2-1},
\end{equation}
where $\tilde{a}_k(x):=\tilde{a}_k(x,x)$. Of course, on using the heat kernel in the form \eqref{heat_kern_res}, we neglect all the exponentially suppressed at $\be\rightarrow0$ terms in the high-temperature expansion. For the complete analytical structure of the heat kernel in the $\tau$-plane see, e.g., \cite{psfss}. The integral over $\tau$ can be easily taken
\begin{equation}\label{int_tau}
    \int_{C}d\tau e^{-\tau (\frac{\omega^2}{\xi^2}-\tilde{m}^2)}\tau^{k-d/2-1}=(-1)^k\frac{2\pi i e^{-i\pi d/2}}{\Gamma(d/2-k+1)}\theta\Big(\frac{\omega^2}{\xi^2}-\tilde{m}^2\Big)\Big(\frac{\omega^2}{\xi^2}-\tilde{m}^2\Big)^{d/2-k}.
\end{equation}
Comparing \eqref{int_tau}, \eqref{om_pot_exp1}, and \eqref{metric}, we see that the imaginary units are all canceled out and the expression \eqref{om_pot_exp1} is real as it should be. The formula \eqref{int_tau} ought to be understood in a distributional sense for $k\geq d/2$. The simplest way to take this into account is to assume that the variable $d$ is complex and tends to its physical value $3$. This rule can be check by convolving \eqref{int_tau} with a test function single-valued near the real positive semiaxis $\omega\geq0$. Only for those test functions does the prescription of the analytical continuation work. In our case this requirement is evidently fulfilled.

Now the integral over $\omega$ has to be evaluated. The heat kernel expansion coefficients $\tilde{a}_k(\omega)$ are polynomial in $\omega$:
\begin{equation}\label{a_k_tild}
    \tilde{a}_k(\omega,x)=:\sum_{j=0}^{2k}\tilde{a}_k^{(j)}(x)(g^2\omega^2)^{j/2},
\end{equation}
where the upper summation limit is dictated by dimensional reasons. Observe that the heat kernel
\begin{equation}\label{heat_kern}
    \lan x|e^{-\tau H_{G}(\omega)}|y\ran
\end{equation}
is Hermitian at the real $\tau$ and therefore its diagonal is real in this case. On the other hand, from Eq. \eqref{H_G} we see that the complex conjugation effectively results in a change of the sign of $\omega$. Hence, by the uniqueness of the analytic continuation in $\tau$, the diagonal of \eqref{heat_kern} is an even function of $\omega$ for any $\tau$. Then the polynomial \eqref{a_k_tild} contains solely the even powers of the energy $\omega$ as long as $E$ is an even function of $\omega$.

So, we need to take the integral
\begin{equation}\label{I}
    I=\int_{0}^{\infty}\frac{\beta d\omega }{e^{\beta\omega}\pm1}\frac{\theta(\omega^2g^2-\tilde{m}^2)}{\Ga(d/2-k+1)}(g^2\omega^2)^{d/2-k+j/2}\Big(1-\frac{\tilde{m}^2}{\omega^2g^2}\Big)^{d/2-k}.
\end{equation}
The further procedure is quite analogous to the case of a flat spacetime (see, e.g., \cite{olopq,DolJack,HabWeld,Kapusta}). We shall follow \cite{olopq}. Substituting the expansion
\begin{equation}
    \Big(1-\frac{m^2}{\omega^2 g^2}\Big)^{d/2-k}=\sum_{n=0}^{\infty}\frac{(-1)^n\Gamma(d/2-k+1)}{n!\Gamma(d/2-k-n+1)}\Big(\frac{m^2}{\omega^2 g^2}\Big)^n
\end{equation}
to the integral $I$, we reduce it to a sum of the incomplete zeta functions. Furthermore, we expand these incomplete zeta functions making use of the formulas \cite{GrRy}:
\begin{equation}\label{incompl_zeta}
\begin{split}
    \int_a^\infty d\omega\frac{\omega^{\nu-1}}{e^\omega-1}&=\Gamma(\nu)\zeta(\nu)-\sum_{n=-1}^\infty\frac{(-1)^n\zeta(-n)a^{\nu+n}}{\Gamma(n+1)(\nu+n)},\\
    \int_a^\infty d\omega\frac{\omega^{\nu-1}}{e^\omega+1}&=(1-2^{1-\nu})\Gamma(\nu)\zeta(\nu)-\sum_{n=0}^\infty(1-2^{1+n})\frac{(-1)^n\zeta(-n)a^{\nu+n}}{\Gamma(n+1)(\nu+n)}.
\end{split}
\end{equation}
The term with $n=-1$ is understood as the limit $n\rightarrow-1$. For the bosonic case we obtain
\begin{multline}\label{I_b}
    I_b=\sum_{n=0}^\infty\frac{(-1)^n\tilde{m}^{d-2k}\be_T^{-j}}{n!\Ga(d/2-k-n+1)}\Big[\Ga(d+j-2k-2n+1)\zeta(d+j-2k-2n+1)(\be_T\tilde{m})^{2k+2n-d}-\\
    -\sum_{l=-1}^\infty\frac{(-1)^l\zeta(-l)(\be_T\tilde{m})^{j+l+1}}{\Ga(l+1)(d+j-2k-2n+l+1)}\Big],
\end{multline}
where $\be_T:=\sqrt{\xi^2}\be$ is the Tolman reciprocal temperature and we have assumed that $\tilde{m}^2\geq0$. The sum over $n$ of the second term in the square brackets can be taken and is expressed through the beta function:
\begin{equation}
    \sum_{n=0}^\infty\frac{(-1)^n}{n!\Ga(d/2-k-n+1)(d+j-2k-2n+l+1)}=-\frac{\Ga(k-(d+j+l+1)/2)}{2\Ga\big((1-j-l)/2\big)}.
\end{equation}
Whence we get
\begin{multline}\label{I_b2}
    I_b=\sum_{n=0}^\infty\frac{(-1)^n\Ga(d+j-2k-2n+1)\zeta(d+j-2k-2n+1)}{n!\Ga(d/2-k-n+1)}\tilde{m}^{2n}\be_T^{2k+2n-j-d}+\\
    +\sum_{l=-1}^\infty\frac{(-1)^l\zeta(-l)\Ga(k-(d+j+l+1)/2)}{2\Ga(l+1)\Ga\big((1-j-l)/2\big)}\tilde{m}^{d-2k+j+l+1}\be_T^{l+1}.
\end{multline}
So, we have all what we need to obtain the high-temperature expansion.

\section{High-temperature expansion}\label{HTE}

Substituting the expression \eqref{I_b2} to \eqref{om_pot_exp1}, we eventually arrive at
\begin{multline}\label{omega_b}
    -F_b=\sum_{k,j=0}^\infty\int d\spx\frac{\sqrt{|g|}\tilde{a}_k^{(j)}}{(4\pi)^{d/2}}\Big[\sum_{n=0}^\infty\frac{(-1)^{n+k}\Ga(d+j-2k-2n+1)\zeta(d+j-2k-2n+1)}{n!\Ga(d/2-k-n+1)\be_T^{d+j-2k-2n+1}\tilde{m}^{-2n}}+\\
    +\sum_{l=-1}^\infty\frac{(-1)^{l+k}\zeta(-l)\Ga(k-(d+j+l+1)/2)}{2\Ga(l+1)\Ga\big((1-j-l)/2\big)}\tilde{m}^{d+j-2k+l+1}\be_T^{l}\Big],
\end{multline}
where the limit of the infinite $\la$ is implied. The expression standing in the square brackets is regular, when $d$ tends to a positive integer, since \eqref{I_b} is the entire function of $d$. Therefore, we can set $d$ to its physical value in the factor at this square brackets. Besides, due to the property of the zeta function a ``half'' of the terms in the sum over $l$ vanish. These are the terms with $l=2,4,6,\ldots$ As far as the fermions are concerned, similar calculations lead to ($\tilde{m}^2\geq0$)
\begin{multline}\label{omega_f}
    -F_f=\sum_{k,j=0}^\infty\int d\spx\frac{\sqrt{|g|}\tilde{a}_k^{(j)}}{(4\pi)^{d/2}}\times\\
    \times\Big[\sum_{n=0}^\infty(1-2^{2n+2k-j-d})\frac{(-1)^{n+k}\Ga(d+j-2k-2n+1)\zeta(d+j-2k-2n+1)}{n!\Ga(d/2-k-n+1)\be_T^{d+j-2k-2n+1}\tilde{m}^{-2n}}+\\
    +\sum_{l=0}^\infty(1-2^{1+l})\frac{(-1)^{l+k}\zeta(-l)\Ga(k-(d+j+l+1)/2)}{2\Ga(l+1)\Ga\big((1-j-l)/2\big)}\tilde{m}^{d+j-2k+l+1}\be_T^{l}\Big].
\end{multline}
The terms at the fixed power of $\be$ both for bosons and fermions are the same up to an overall numeric factor and in that sense are universal. However, the expansion of the free energy for bosons involves one additional contribution with $l=-1$.

Let us analyze the first and second terms in the square brackets of \eqref{omega_b} and \eqref{omega_f} separately. We see that both for bosons and fermions the second terms in the square brackets are not expanded in the increasing powers of $\beta$. Rather, we have a derivative (or large mass) expansion of the terms at the every fixed power of the reciprocal temperature. It is not difficult to write out a closed form expression for a whole series at the fixed power of $\beta$. With this end in view, we introduce the function (cf. \cite{olopq,DowKen})
\begin{equation}\label{sigma}
    \s^l_\nu(m^2):=\int_C\frac{ds s^{\nu-1}}{(e^{2\pi i\nu}-1)\Ga(\nu)}\int_0^\infty d\omega\omega^l\Tr_d e^{-sH(\omega)}.
\end{equation}
The function under the integral over $s$ given by the integral over $\omega$ is analytic for $\re s>0$. It should be analytically continued to the imaginary axis and to the vicinity of the origin where the contour $C$ lies. Usually, this can be achieved by rotating the integration contour in the $\omega$-plane (see the high frequency asymptotics in Eq. \eqref{om_pot_exp1}). Notice that the function \eqref{sigma} is not the generalized zeta function as long as $H(\omega)$ possesses the negative eigenvalues. Although these functions are closely related. Repeating all the above calculations, one can convince oneself that the expansion \eqref{omega_b} can be rewritten as
\begin{multline}\label{high_temp}
    -F_b=\sum_{k,n,j=0}^\infty\int d\spx\frac{\sqrt{|g|}\tilde{a}_k^{(j)}}{(4\pi)^{d/2}}\frac{(-1)^{k}\Ga(d+j-2k-2n+1)\zeta(d+j-2k-2n+1)}{n!\Ga(d/2-k-n+1)\be_T^{d+j-2k-2n+1}(-\tilde{m}^2)^{-n}}+\\
    +\sum_{l=-1}^\infty\frac{(-1)^{l}\zeta(-l)}{\Ga(l+1)}\s_\epsilon^l\be^{l},
\end{multline}
where $\epsilon=(\bar{d}-d)/2$ is the complex number tending to zero and $\bar{d}$ is the physical dimension ($\bar{d}=3$). The analogous formula holds for fermions as well. The function $\s_\nu^l(m^2)$ depends only on $m^2$ and the background fields. Consequently, we need certain additional assumptions apart from $\be\rightarrow0$ in order to obtain an explicit expression for it. Henceforth we assume that the gravitational field varies slowly such that the derivative expansions presented in \eqref{omega_b} and \eqref{omega_f} make sense and provide reliable approximations for the functions \eqref{sigma}. This holds when the characteristic length of the variation of the gravitational field is much larger than the Compton wavelength associated with the effective mass $\tilde{m}$. In any case, these second terms give subleading contributions to the high-temperature expansion. For fermions, they stand at the nonegative integer powers of $\beta$, while for bosons the leading contribution from these terms is proportional to $\be^{-1}$. The first terms in the square brackets in the expansions \eqref{omega_b} and \eqref{omega_f} are much more singular at $\be\rightarrow0$.

Consider the first terms in the square brackets of Eqs. \eqref{omega_b} and \eqref{omega_f}. There is a finite number of such terms at any fixed power of $\be$. Indeed, at any fixed number $k$, the number $j$ must be less or equal to $2k$ by dimensional reasons. However, inasmuch as we resummed the expansion and collected all the terms without derivatives of $E$ to the exponent, the latter terms are absent in $\tilde{a}_k$. Consequently, the worst terms, which contain the maximal power of $\omega$ at the fixed dimension $2k$, are of the form
\begin{equation}\label{nablaE}
    \nabla E\nabla E\cdots\nabla E.
\end{equation}
If the dimension $2k$ is not a multiple of $3$ then the worst terms look like \eqref{nablaE}, but with the additional covariant derivative, or with $\Omega^2$ (see, for example,  \eqref{a_5}). Thus, we conclude that (see also \cite{NakFuk,BarvMukh})
\begin{equation}
    j\leq[4k/3],
\end{equation}
where the square brackets denote the integer part of the number. These worst terms stand at
\begin{equation}\label{cond_k}
    \be^{2n+2k-[4k/3]-d-1},
\end{equation}
and so the power of $\be$ increases. Further, we restrict ourself by the singular and finite parts of the high-temperature expansion as they give the leading contribution to the free energy. From \eqref{cond_k} we conclude that the maximal number $k$ that we need is determined by the equality
\begin{equation}
    2k-[4k/3]=D,
\end{equation}
and for a three dimensional space $k\leq6$. Fortunately, the heat kernel coefficients $a_k$ with $k\leq6$ were found in \cite{Ven} and we just can borrow these results. The relevant parts of these coefficients are presented in Appendix \ref{HKE_coeff}.

Now we turn again to the second terms in the square brackets of the expansions \eqref{omega_b} and \eqref{omega_f}. The contributions with the positive even $l$ vanish. Moreover, the contribution from these terms is zero when $l$ is an odd positive number and $j$ is an even nonegative number. As we discussed above, the number $j$ must be even and nonegative. The explicit expressions for $\tilde{a}_k$ given in Appendix \ref{HKE_coeff} are found to be even functions of $\omega$ for those $\tilde{a}_k$ which we are interested in and confirm thereby the general considerations.

Note that these second terms were disregarded in \cite{Fursaev1,Fursaev2,NakFuk}. It is relevant for a correct evaluation of these terms that we did not make a conformal transformation to the optic metric and resummed the heat kernel expansion. The resummation gives a non-perturbative expression for the heat kernel which is more reliable in terms of the original metric rather than the optic one (see for details \cite{psfss}). In principle, the first contributions in the square brackets of Eqs. \eqref{omega_b} and \eqref{omega_f} may be evaluated in terms of the conformally transformed metric. However, in order to provide a transparent way to the poles cancelation in the $d$-plane they have to be expressed through the same metric as the second terms.

\section{Results and implications}\label{ResImpl}

Bearing in mind these observations, we can write for $d=3$ in the bosonic case
\begin{multline}\label{F_b}
    -F_b=\int \frac{d\spx\sqrt{|g|}}{(4\pi)^{d/2}}\sum_{k,j=0}^\infty(-1)^k\Big\{\sideset{}{'}\sum_{n=0}^\infty\frac{\Ga(D+j-2k-2n)\zeta(D+j-2k-2n) \tilde{a}_k^{(j)}}{n!\Ga(d/2-k-n+1)\be_T^{D+j-2k-2n}(-\tilde{m}^2)^{-n}}\\
    +\frac{(-\tilde{m}^2)^{(D+j)/2-k}\tilde{a}_k^{(j)}}{4((D+j)/2-k)!\Ga((1-j)/2)} \big[\ln\frac{\tilde{m}^2\be_T^2e^{2\ga}}{4\pi^2}-\psi((D+j)/2-k+1)+\psi((1-j)/2)\big]\\
    +\de_{j,0}\Ga(k-d/2)\frac{\tilde{a}_k^{\br0}}{2\be_T}\tilde{m}^{d-2k}-\tilde{a}_k^{(j)}\tilde{m}^{D+j-2k}\frac{\Ga(k-(D+j)/2)}{4\Ga((1-j)/2)}\Big\},
\end{multline}
where $\ga$ is the Euler constant. The prime at the sum over $n$ says that the singular terms are discarded, the second term is zero by definition when the argument of the factorial is negative, and the last term vanishes by definition when the gamma function entering the numerator tends to infinity. The last term stands at the negative power of $\tilde{m}$. Later on we shall cast out such contributions although it is these contributions which are ``protected'' from the high energy physics, i.e., the particles with a large mass (but with $m\ll\beta^{-1}$) give a negligible contribution to these terms. The high-temperature expansion in the fermionic case looks like
\begin{multline}\label{F_f}
    -F_f=\int \frac{d\spx\sqrt{|g|}}{(4\pi)^{d/2}}\sum_{k,j=0}^\infty(-1)^k\Big\{\sideset{}{'}\sum_{n=0}^\infty(1-2^{2k+2n-j-d})\frac{\Ga(D+j-2k-2n)\zeta(D+j-2k-2n) \tilde{a}_k^{(j)}}{n!\Ga(d/2-k-n+1)\be_T^{D+j-2k-2n}(-\tilde{m}^2)^{-n}}\\
    -\frac{(-\tilde{m}^2)^{(D+j)/2-k}\tilde{a}_k^{(j)}}{4((D+j)/2-k)!\Ga((1-j)/2)} \big[\ln\frac{4\tilde{m}^2\be_T^2e^{2\ga}}{\pi^2}-\psi((D+j)/2-k+1)+\psi((1-j)/2)\big]\\
    +\tilde{a}_k^{(j)}\tilde{m}^{D+j-2k}\frac{\Ga(k-(D+j)/2)}{4\Ga((1-j)/2)}\Big\}.
\end{multline}
The digamma functions appearing in the expressions are easily calculated using the formulas \cite{GrRy}:
\begin{equation}
    \psi\Big(\frac12-n\Big)=\sum_{k=1}^n\frac{2}{2k-1}-\ln4-\ga,\qquad\psi(1+n)=\sum_{k=1}^n\frac1{k}-\ga.
\end{equation}
More specifically, taking into account that $\tilde{a}_0=1$ and $\tilde{a}_1=0$, we derive for bosons
\begin{equation}\label{omega_b_fin}
\begin{array}{l}
    -F_b=\int d\spx\sqrt{|g|}\Big\{\dfrac{\pi^2\beta^{-4}_T}{90} -\dfrac{\be^{-2}_T}{24}(\tilde{m}^2-\frac12\tilde{a}_2^{\br2}-\frac14\tilde{a}_3^{\br4}) +\dfrac{\be_T^{-1}}{12\pi}(\tilde{m}^3+\frac3{4\tilde{m}}\tilde{a}_2^{\br0}+\ldots)\\
    +\dfrac1{64\pi^2}\ln\dfrac{\tilde{m}^2\be_T^2e^{2\ga}}{16\pi^2}\big[\tilde{m}^4+\tilde{m}^2(\tilde{a}_2^{\br2}+\frac32 \tilde{a}_3^{\br4})+2\tilde{a}_2^{\br0} +\tilde{a}_3^{\br2}+\frac32\tilde{a}_4^{\br4}+\frac{15}4\tilde{a}_5^{\br6}+\frac{105}8\tilde{a}_6^{\br8}\big]\\[1em]
    -\dfrac{3}{128\pi^2}\big[\tilde{m}^4-\tilde{m}^2(\frac23\tilde{a}_2^{\br2}+\frac53\tilde{a}_3^{\br4})-\frac43(\tilde{a}_3^{\br2}+2\tilde{a}_4^{\br4}+\frac{23}4\tilde{a}_5^{\br6}+22\tilde{a}_6^{\br8}) \big]+\ldots
    \Big\},
\end{array}
\end{equation}
and for fermions
\begin{equation}\label{omega_f_fin}
\begin{array}{l}
    -F_f=\int d\spx\sqrt{|g|}\Big\{\dfrac{7\pi^2\beta^{-4}_T}{720} -\dfrac{\be^{-2}_T}{48}(\tilde{m}^2-\frac12\tilde{a}_2^{\br2}-\frac14\tilde{a}_3^{\br4})\\
    -\dfrac1{64\pi^2}\ln\dfrac{\tilde{m}^2\be_T^2e^{2\ga}}{\pi^2}\big[\tilde{m}^4+\tilde{m}^2(\tilde{a}_2^{\br2}+\frac32 \tilde{a}_3^{\br4})+2\tilde{a}_2^{\br0} +\tilde{a}_3^{\br2}+\frac32\tilde{a}_4^{\br4}+\frac{15}4\tilde{a}_5^{\br6}+\frac{105}8\tilde{a}_6^{\br8}\big]\\[1em]
    +\dfrac{3}{128\pi^2}\big[\tilde{m}^4-\tilde{m}^2(\frac23\tilde{a}_2^{\br2}+\frac53\tilde{a}_3^{\br4})-\frac43(\tilde{a}_3^{\br2}+2\tilde{a}_4^{\br4}+\frac{23}4\tilde{a}_5^{\br6}+22\tilde{a}_6^{\br8}) \big]+\ldots
    \Big\}.
\end{array}
\end{equation}
The dots in the expressions above denote the terms at the negative powers of $\tilde{m}$ or the positive powers of $\beta$. The coefficients diverging at $\be\rightarrow0$ provide the energy-time anomaly (\cite{psfss} and also \cite{BrOtPa,FrZel}) when the energy cutoff is used for the regularization of the one-loop contribution to the effective action. Recall that the energy-time anomaly represents the variance of the renormalized effective action under the uniform dilatations of the Killing vector (see below). The flat spacetime limit of the expansions \eqref{omega_b_fin} and \eqref{omega_f_fin} coincides with the known one \cite{DolJack,Kapusta,HabWeld}.

Now we consider the coefficients at the different powers of $\be$ in detail. First, we observe that
\begin{equation}\label{T_2}
    \int d\spx\sqrt{|g|}\be^{-2}_T(\tilde{m}^2-\frac12\tilde{a}_2^{\br2}-\frac14\tilde{a}_3^{\br4})=\be^{-2}\int d\spx\sqrt{|g|}\big[(m^2+\frac16 R)\xi^{-2}+\frac1{12}f^2-\frac13\nabla^\mu(g^2h_\mu)\big].
\end{equation}
If we omit the last term representing a total divergence then we confirm the results of \cite{DowKen,NakFuk,DowSch} that the term at $\be^{-2}$ is absent for a massless conformal scalar field on a static gravitational background. Meanwhile, we see that this is not the case for a stationary spacetime. For such spacetimes, the leading correction to the Planck law is of the order of $\be^{-2}$ and proportional to the Maxwell-like term $f^2$. Formula \eqref{T_2} coincides with Eq. (4.18) of \cite{Fursaev1} up to the integral of a total derivative. Under certain circumstances the last term in \eqref{T_2} reducing to a boundary term also considerably contributes. However, in this case other boundary contributions that we discarded earlier must be included into the free energy too.

In the bosonic case, the next correction is proportional to $\be^{-1}$. For a massless scalar field on a curved background satisfying the vacuum Einstein equations $R_{\mu\nu}=0$, this contribution becomes purely imaginary as long as the correction to $m^2$ is negative in this case \cite{psfss}. There is also an imaginary contribution to the free energy due to the logarithmic term. The imaginary contributions to the effective action signalize about a certain instability of the vacuum of quantum fields which was initially chosen. As a result, we may anticipate that a non-zero vacuum expectation value of the scalar field develops (the Bose-Einstein condensation).

Of course, if we accurately evaluate the high-temperature expansion of \eqref{om_potent} then there are not any imaginary contributions to the free energy as the expression \eqref{om_potent} is real-valued. For $\tilde{m}^2<0$, the additional terms appear in \eqref{omega_b_fin} and \eqref{omega_f_fin} such that the imaginary part of the free energy vanishes. In the fermionic case, the contributions to the real part of the free energy standing at $\be^{-4}$, $\be^{-2}$, $\be^0$, and $\ln\be$ do not change, i.e., the additional terms cancel exactly the imaginary part coming from the logarithm and add certain contributions at the higher powers of $\be$. At the same time, in the bosonic case, the imaginary part is also canceled out, but the infrared divergence arises in the term with $\be^{-1}$. Other contributions to the singular and finite parts of the high-temperature expansion remain unchanged. The term at $\be^{-1}$ entering the free energy $F_b$ becomes
\begin{multline}
%\begin{array}{l}
    \int d\spx\frac{\sqrt{|g|}}{(4\pi)^{d/2}}\sum_{k=0}^\infty\frac{(-1)^k}{2\be_T}\bigg\{\frac{|\tilde{m}|^{d-2k}\tilde{a}_k^{\br0}}{\Ga(d/2-k+1)}\big[\ln\frac{\bar{\la}^2}{\xi^2|\tilde{m}|^2}+\ga+\psi(d/2-k+1) \big]-\\
    -\sum_{j=1}^{[4k/3]} \tilde{a}_k^{(j)}(-1)^{j/2}\frac{|\tilde{m}|^{d+j-2k}\Ga(j/2)}{\Ga((d+j)/2-k+1)} \bigg\}
    =\int \frac{d\spx\sqrt{|g|}}{12\pi^2\be_T}\Big[|\tilde{m}|^3\ln\frac{g^2\bar{\la}^2e^{8/3}}{4|\tilde{m}|^2}+\frac{3|\tilde{m}|}{2}(\tilde{a}^{(2)}_2+ \tilde{a}^{(4)}_3)+\ldots \Big],
%\end{array}
\end{multline}
where $\bar{\la}$ is the infrared energy cutoff. The contributions vanishing at $\bar{\la}\rightarrow0$ were neglected and dots denote the negative powers of $\tilde{m}$.

These corrections to \eqref{omega_b_fin} and \eqref{omega_f_fin} have their origin in the fact that at $\tilde{m}^2<0$ the integral \eqref{I} can be represented as the sum of two integrals along the contours $[\bar{\la},i\sqrt{\xi^2}|\tilde{m}|]$ and $[i\sqrt{\xi^2}|\tilde{m}|,+\infty)$. The latter integral was already evaluated and its contribution to the high-temperature expansion is given by \eqref{F_b} or \eqref{F_f}. The former integral,
\begin{equation}
    \int_{\bar{\la}}^{i\sqrt{\xi^2}|\tilde{m}|}\frac{\be d\omega}{e^{\be\omega}\pm1}\frac{(g^2\omega^2)^{j/2}(g^2\omega^2-\tilde{m}^2)^{d/2-k}}{\Ga(d/2-k+1)},
\end{equation}
is readily taken if one expands $(e^{\be\omega}\pm1)^{-1}$ in a Laurent series in $\be\omega$. The infrared cutoff is only needed in the bosonic case for the contribution at $(\be\omega)^{-1}$ and $j=0$:
\begin{equation}
    \int_{\bar{\la}}^{i\sqrt{\xi^2}|\tilde{m}|}\frac{\be d\omega}{e^{\be\omega}-1}\frac{(g^2\omega^2-\tilde{m}^2)^{d/2-k}}{\Ga(d/2-k+1)}\approx\int_{\bar{\la}}^{i\sqrt{\xi^2}|\tilde{m}|}\frac{ d\omega}{\omega}\frac{(g^2\omega^2-\tilde{m}^2)^{d/2-k}}{\Ga(d/2-k+1)}.
\end{equation}
This integral, in turn, can be estimated at $\bar{\la}\rightarrow0$ making use of the expansion of the incomplete beta function \cite{GrRy}:
\begin{equation}
\begin{split}
    \int_a^1dx x^{\al-1}(1-x)^{\be-1}&=\frac{\Ga(\al)\Ga(\be)}{\Ga(\al+\be)}-\sum_{n=0}^\infty\frac{(-1)^n\Ga(\be)a^{n+\al}}{n!\Ga(\be-n)(n+\al)},\\
    \int_a^1\frac{dx}{x}(1-x)^{\be-1}&=-\ga-\psi(\be)-\ln a-\sum_{n=1}^\infty\frac{(-1)^n\Ga(\be)a^{n}}{n!\Ga(\be-n)n}.
\end{split}
\end{equation}

To sum up, we see that at $\tilde m^2\leq0$ the infrared divergencies arise in the free energy. Therefore, a more accurate analysis of the low energy modes is required. These divergencies may be cured by introducing the Bose-Einstein condensate and summing the ring diagrams (see, e.g., \cite{Kapusta,DolJack}) or, as in the flat spacetime, by introducing the boundaries and accounting for the finite size effects. In the case, when
\begin{equation}
    l\lesssim m^{-1}\ll L,
\end{equation}
where $l$ is the characteristic scale of the variations of the gravitational field and $L$ is the size of the ``box'', the boundary effects are negligible and so the condensation is the only possible mean to remove the infrared divergencies. Keeping in mind that the free energy obtained above is the finite-temperature part of the effective action, the appearance of the infrared divergencies at $m^2\rightarrow0$ for a slowly varying gravitational field is quite expectable (see, e.g., \cite{Kapusta,DolJack,Weinberg,GrPeYa,Akhmed,Polyak}). For the rapidly changing metric field such divergencies may not appear \cite{GusZeln}, but this fact has nothing in common with the infrared divergencies arising in the infrared limit.

The coefficient of the logarithmic term is closely related with the conformal (trace) anomaly \cite{DowKen,DowSch}. Substituting the expressions for $\tilde{a}_k^{(j)}$ from Appendix \ref{HKE_coeff}, we come to  \cite{DowKen,DowSch,NakFuk,GusZeln,Fursaev2}
\begin{equation}\label{e-t_anom}
\begin{array}{l}
    \tilde{m}^4+\tilde{m}^2(\tilde{a}_2^{\br2}+\frac32 \tilde{a}_3^{\br4})+2\tilde{a}_2^{\br0} +\tilde{a}_3^{\br2}+\frac32\tilde{a}_4^{\br4}+\frac{15}4\tilde{a}_5^{\br6}+\frac{105}8\tilde{a}_6^{\br8}=\\
    =m^4+\frac13m^2R+\frac1{90}(R^{\mu\nu\rho\s}R_{\mu\nu\rho\s} -R^{\mu\nu}R_{\mu\nu}+6\nabla^2R)+\frac1{36}R^2.
\end{array}
\end{equation}
A great number of cancelations happens in calculating this expression. This fact can be considered as the indirect verification of the correctness of the coefficients entering the expansion above. Remarkably, this expression does not depend on the Killing vector field. The nonminimal coupling adds to the anomaly \eqref{e-t_anom} the term,
\begin{equation}
    -2m^2\kappa R+(\kappa^2-\frac{\kappa}3)R^2-\frac{\kappa}{6}\nabla^2R,
\end{equation}
as one can easily deduce from the substitution rule \eqref{nonmin}. In the conformal case $m^2=0$, $\kappa=1/6$, the coefficient of the logarithm coincides with the standard expression for the conformal anomaly.

Some comments on the conformal transformations are in order to caution the interested reader. In the paper \cite{DowSch}, the high-temperature expansion of the free energy of a scalar field on a static gravitational background was obtained by the use of the conformal transform from the ultrastatic spacetime. Despite the fact that the final result  (Eq. (18), \cite{DowSch}) for the terms at $\be^{-4}$, $\be^{-2}$, $\be^{-1}$, and $\ln\be$ is correct for the static gravitational field, the method for its derivation seems to be invalid at the point where the scaling property of the generalized zeta function (the equation before Eq. (10) of \cite{DowSch}) was employed. This scaling property holds only for the conformally invariant operator ($m^2=0$ and $\kappa_D=(D-2)/4(D-1)$) and does not take place in a general case (see, e.g., \cite{DowHKEscal}). Of course, one can always make a transformation
\begin{equation}\label{conf_trans}
\begin{array}{l}
    g_{\mu\nu}\rightarrow\bar{g}_{\mu\nu}=e^{2\rho}g_{\mu\nu},\\
    |\bar{g}|^{1/4}\bar{\nabla}^2|\bar{g}|^{-1/4}-\kappa\bar{R}+m^2=\\
    =e^{-\rho}\Big[|g|^{1/4}\nabla^2|g|^{-1/4}-\kappa R+(D-1)(\kappa-\kappa_D)\big(2\nabla^2\rho +(D-2)\nabla_\mu\rho\nabla^\mu\rho\big) +m^2e^{2\rho}\Big]e^{-\rho},
\end{array}
\end{equation}
with $\rho$ independent of time. Nevertheless, the variation of this expression with respect to $\rho$ gives not only the term arising in the scaling property of the zeta function \cite{DowSch}, but also the contribution from the variation of the expression standing in the square brackets \eqref{conf_trans}. This contribution is not trivial in the case $\kappa\neq\kappa_D$ and $m^2\neq0$.

%!or at least to impel him/her to draw attention to the following point!

On the other hand \cite{DowSch1}, one can indeed make the substitution \eqref{conf_trans} from the beginning \eqref{KG_eq} and apply the whole machinery of the heat kernel expansion \eqref{HKE_ini} to the operator standing in the square brackets. The spectrum associated with this operator is obviously the same as the spectrum associated with the initial operator, and so the high-temperature expansions must coincide. If the function $\rho$ is of a general type, such that, in particular, the coefficients at $\omega^2$ and $m^2$ in $H(\omega)$ are not constants, then this function disappears from the high-temperature expansion upon restoration of the dependence $g_{\mu\nu}$ on $\rho$. However, for a certain particular choice of $\rho$ yielding a vanishing of some terms in the coefficients $a_k(\omega,x)$ of the heat kernel expansion, for example, for $\rho=\ln\sqrt{\xi^2}$ leading to the constant coefficient at $\omega^2$ in $H(\omega)$, to the optical metric, and to the needlessness of the resummation of the expansion for a static metric \cite{DowSch,DowSch1}, the dependence of the free energy on this $\rho$ does not disappear, in general. In making the conformal transformations, one should take into account that the coefficients $a^{(j)}_k$, $j\geq2$, become nonvanishing and they also contribute to the free energy or, in other words, the additional terms arise in the scaling relation for the generalized zeta function \cite{DowHKEscal}. This, of course, is just a manifestation of the fact that the Klein-Gordon equation is not conformally invariant for the arbitrary $\kappa$ and $m^2$.

Now we return to the discussion of the properties of the expansions \eqref{omega_b_fin} and \eqref{omega_f_fin}. The contribution to the finite part of the free energy at $\be\rightarrow0$, which determines, in particular, the Casimir force at zero temperature, does depend on the Killing vector. The whole expression is rather huge, and we do not write it here (see Eq. \eqref{fin_part}), but the terms at $m^2$ take a compact form:
\begin{equation}
    \tilde{m}^4-\tilde{m}^2(\frac23\tilde{a}_2^{\br2}+\frac53\tilde{a}_3^{\br4})+\ldots=m^4+\frac{m^2}3(R+\frac53\xi^2R_{\mu\nu}g^\mu g^\nu+\frac{11}{6}h^2)+\ldots
\end{equation}
These terms give the leading contribution to the finite part when $m^{-1}$ is much smaller than the characteristic scale of variations of the gravitational field. The finite part does not coincide with the answers presented in \cite{DowSch} and \cite{DowSch1} (compare the coefficient at $h^2$). Recalling that the masses of all the massive particles are generated by the Higgs mechanism, we see that the term at $m^2$ gives a contribution to the effective potential of the Higgs field. This results in the additional correction to the gravitational mass-shift effect discussed in \cite{gmse}.

If one regards the expression \eqref{omega_f_fin} as the regularized version of the one-loop contribution to the effective action at zero temperature with $\be$ playing the role of the regularization parameter (see Eq. \eqref{one_loop_bos}), then an interesting relation between the conformal and energy-time anomalies can be observed. Recall that the corresponding global symmetries lead to the (formal) Noether theorem
\begin{equation}\label{anomalies}
    g_{\mu\nu}(x)\frac{\de\Ga}{\de g_{\mu\nu}(x)}=2\sqrt{|g|}\nabla_\mu D^\mu,\qquad \xi^\mu(x)\frac{\de\Ga}{\de\xi^\mu(x)}=\sqrt{|g|}\nabla_\mu T^\mu,
\end{equation}
where $\Ga[g_{\mu\nu},\xi^\mu,\Phi]$ is the renormalized effective action, and $D^\mu$ and $T^\mu$ are the corresponding currents. Usually, these relations are broken by the anomaly terms resulting from the quantum corrections. In our particular case (the massless conformal scalar field), we can renormalize the regularized contribution \eqref{omega_f_fin} to the effective action in several ways.

The first one: we add the following counterterms to the initial Lagrangian
\begin{equation}\label{c-t_1}
    \be^{-4}(g_\mu g^\mu)^2,\qquad \be^{-2}f^2,\qquad [\tilde{m}^4+\tilde{m}^2(\tilde{a}_2^{\br2}+\frac32 \tilde{a}_3^{\br4})+\cdots]\ln(\be^{-2}g^2).
\end{equation}
As for the last counterterm, such a choice is unnatural and would be prohibited if the quantum theory of the vector field $g_\mu$ will be renormalizable (cf. the Coleman-Weinberg potential \cite{ColWein} containing the term $\phi^4\ln\phi^2$). Nevertheless, if we add such counterterms, the global conformal invariance of the renormalized effective action is broken as long as this logarithmic counterterm is not conformally invariant. The conformal anomaly is given by the factor at the logarithm. On the other hand, the second relation in \eqref{anomalies} holds on a quantum level provided the power-like divergencies are completely canceled out. The finite part and the coefficient of the logarithm are invariant under the stretching of the Killing vector.

The second way: we add to the initial Lagrangian the counterterms
\begin{equation}\label{c-t_2}
    \be^{-4}(g_\mu g^\mu)^2,\qquad \be^{-2}f^2,\qquad [\tilde{m}^4+\tilde{m}^2(\tilde{a}_2^{\br2}+\frac32 \tilde{a}_3^{\br4})+\cdots]\ln(\be^{-2}),
\end{equation}
the coefficients at the power-like divergencies being renormalized to finite constants (not necessarily to zero). Then the second relation in \eqref{anomalies} is broken, while the first one is left untouched \cite{DeWQFTcspt}. All the counterterms are invariant even under the local conformal transformations up to total derivatives.

So, we face with a choice what the symmetry we want to preserve on a quantum level. Whatever the case, the dependence of the effective action on the Killing vector cannot be completely removed. The imaginary part of the effective action, which is responsible for the Hawking particle production \cite{Hawk}, depends on the Killing vector.

The field $g_\mu$ plays a distinguished role in defining the state of quantum fields. The structure of the counterterms suggests the simplest form of the classical action for this field. Its Lagrangian density reads \cite{gmse,KostelBB,Will,DeWQFTcspt}
\begin{equation}\label{action}
    \mathcal{L}=-\frac14f^2-\frac12(\bar{\al}|\phi|^2+\bar{\be}+\bar{\kappa} R)g^2-\frac{\bar{\ga}}{4}g^4,\qquad \bar{\ga}>0,
\end{equation}
where $\phi$ is the Higgs field ($SU(2)$ doublet), $\bar{\al}$, $\bar{\ga}$, and $\bar{\kappa}$ are the dimensionless constants, and $\bar{\be}$ has the mass dimension $2$. This action resembles the simplest model of a ferromagnetic \cite{LandLifshECM} near the Curie point in the Landau theory of phase transitions. The constants $\bar{\al}$ and $\bar{\be}$ are such that the potential possesses the minimum at $g^2>0$. It is natural to suppose that the characteristic magnitude of the vacuum expectation value of the field $g_\mu$ is of the order of the Planck scale. From the renormalization theory viewpoint, it is also appealing to suppose that $\bar{\al} g^2=\mu^2$, where $\mu^2$ is the parameter of the Higgs potential (see \cite{gmse} for the notation). In this case, the constant $\bar{\be}$ can be put to zero and all the coupling constants of the model become dimensionless, apart from the Newton and the cosmological constants. In particular, we have the estimations for the constants determining the potential of the field $g_\mu$: $\bar{\al}\approx \eta^2/m_{Pl}^2$ and $\bar{\ga}\approx\bar{\al}^2$, where $\eta$ is the vacuum expectation value of the Higgs field. It should be noted that the dynamics of the field $g_\mu$ are determined not from the least action principle, but from the self-consistency requirement \cite{gmse} imposed on the effective action. The field $g_\mu$ is at the minimum of the effective potential only for the Minkowski spacetime (see \cite{gmse}). The exploration of the viability of the above model and its phenomenological consequences will be given elsewhere. Also note that the fields with higher spins can induce additional terms in \eqref{action}. The spinor fields, for example, add the contribution proportional to $h^2$ to the terms standing in round brackets in \eqref{action}.

\section{Conclusion}

In this paper, we derived the high-temperature expansion of the free energy in a complete form \eqref{high_temp}. The main results of the paper are collected and discussed in Sec. \ref{ResImpl}. In particular, the explicit expressions for the divergent and finite parts of the high-temperature expansions are given in Eqs. \eqref{omega_b_fin}, \eqref{omega_f_fin} in the case of the four-dimensional spacetime. We generalized the results of the papers \cite{Fursaev1,Fursaev2} concerning the high-temperature expansion of the free energy on stationary spacetimes and found the additional contributions to the high-temperature expansion that were disregarded in those papers. In that part of the expansion which was presented in \cite{Fursaev1,Fursaev2} our expression for the free energy coincides with given in \cite{Fursaev1,Fursaev2} up to integrals of the total derivatives. The method used in the present paper is straightforwardly generalized to the case of the higher spin fields, to the fields with the dispersion law differing from the relativistic one, and to the grand canonical ensembles.

\appendix

\section{Useful relations with the Killing vector}\label{4_Kill}

In this appendix, we collect some formulas regarding the differential calculus on the Riemannian manifold with the Killing vector. Let us introduce the notation:
\begin{equation}
    g_\mu=\frac{\xi_\mu}{\xi^2},\qquad h_\mu=\partial_\mu\ln\sqrt{\xi^2},\qquad f_{\mu\nu}=\partial_{[\mu}g_{\nu]},
\end{equation}
where $\xi^\mu$ is the time-like Killing vector of the metric $g_{\mu\nu}$. Then the following useful relations hold
\begin{equation}
\begin{gathered}
    f_{\mu\la}g^\la=0,\qquad
    g_\la h^\la=0,\qquad
    \nabla_\mu g_\nu=\frac12 f_{\mu\nu}-h_{(\mu}g_{\nu)},\\
    \nabla_\mu h_\nu g^\nu=-\frac12 f_{\mu\nu}h^\nu+g_\mu h^2,\qquad
    g^{\mu}g^{\nu}\nabla_{\mu}h_{\nu}=g^2h^2,\\
    g^\la \nabla_\la f_{\mu\nu}=-g_{[\mu}f_{\nu]\la}h^\la,\qquad
    \nabla_\mu f_{\nu\la}g^\la=\frac12 f^2_{\mu\nu}+g_\mu f_{\nu\la}h^\la,\qquad
    \nabla^\la f_{\la\mu}g^\mu=-\frac12f^2,
\end{gathered}
\end{equation}
and for the curvature
\begin{equation}
\begin{gathered}
    g^\la R_{\la\nu\s\mu}=\frac12\nabla_\nu f_{\s\mu}-\frac12h_{[\s}f_{\mu]\nu}+h_{[\s}g_{\mu]}h_\nu+f_{\s\mu}h_\nu-g_{[\s}\nabla_{\mu]}h_\nu,\\
    g^\la R_{\la\mu}=f_{\mu\la}h^\la-\frac12\nabla^\la f_{\la\mu}-g_\mu\nabla_\la h^\la,\qquad
    R_{\mu\nu}g^\mu g^\nu=\frac14 f^2-g^2\nabla_\la h^\la,
\end{gathered}
\end{equation}
where $f^2:=f_{\mu\nu}f^{\mu\nu}=-\Tr f^2$.

In the course of the derivation of the high-temperature expansion, it was useful to introduce the negative definite metric $G_{\mu\nu}$ (see Eq. \eqref{metric}). The corresponding Christoffel symbols are
\begin{equation}\label{connection}
    \tilde{\Ga}^\rho_{\mu\nu}=\Ga^\rho_{\mu\nu}+\frac12(\la^{-1}-\xi^2)g_{(\mu}f_{\nu)}^{\ \rho}-\xi^2(h_{(\mu}g_{\nu)}g^\rho-g_\mu g_\nu h^\rho)=:\Ga^\rho_{\mu\nu}+\ga^\rho_{\mu\nu},
\end{equation}
where $\Ga^\rho_{\mu\nu}$ are the Christoffel symbols for the metric $g_{\mu\nu}$. Hereinafter all the quantities associated to the metric $G_{\mu\nu}$ are designated by tildas. From \eqref{connection} we see that the connection $\tilde{\Ga}^\rho_{\mu\nu}$ tends to the projected connection (see, e.g., \cite{psfss,LandLifshCTF,Fursaev1,Fursaev2}) when $\la$ goes to infinity. The curvature tensor for the connection $\tilde{\Ga}^\rho_{\mu\nu}$ is defined in the standard way
\begin{equation}
    \tilde{R}^\rho_{\ \nu\s\mu}=R^\rho_{\ \nu\s\mu}+\nabla_{[\s}\ga_{\mu]\nu}^\rho+\ga_{[\s\la}^\rho\ga_{\mu]\nu}^\la.
\end{equation}
After tedious calculations, we arrive at
\begin{multline}
    \tilde{R}^\rho_{\ \nu\s\mu}=R^\rho_{\ \nu\s\mu}+\frac12(\la^{-1}-\xi^2)\Bigl(f_{\s\mu}f_\nu^{\ \rho}+g_{[\mu}\nabla_{\s]}f_{\nu}^{\ \rho}-g_\nu\nabla^\rho f_{\s\mu}-\frac12 f_{\nu[\s}f_{\mu]}^{\ \rho} \Bigr)\\
    -\xi^2\Bigl(g_{[\mu}h_{\s]}h_\nu g^\rho-g_{[\mu}h_{\s]}h^\rho g_\nu+\frac12h_{[\mu}f_{\s]\nu}g^\rho-\frac12g_{[\mu}f_{\s]\nu}h^\rho+g_{[\mu}\nabla_{\s]}h_\nu g^\rho-g_{[\mu}\nabla_{\s]}h^\rho g_\nu+f_{\s\mu}(h_\nu g^\rho-h^\rho g_\nu)+\\
    +f_\nu^{\ \rho} h_{[\s}g_{\mu]}-\frac12 h_{[\mu}f_{\s]}^{\ \rho}g_\nu+\frac12 g_{[\mu}f_{\s]}^{\ \rho}h_\nu \Bigr)+\frac14(\la^{-1}-\xi^2)^2 g_\nu g_{[\s}f_{\mu]}^{\ \la}f_{\la}^{\ \rho}-\frac12(\la^{-1}-\xi^2)\xi^2g_{[\s}f_{\mu]}^{\ \la}h_\la g^\rho g_\nu.
\end{multline}
The Ricci curvature becomes
\begin{multline}
    \tilde{R}_{\mu\nu}=R_{\mu\nu}+\frac{1}{2}(\la^{-1}-\xi^2)[f^2_{\mu\nu}-g_{(\mu}\nabla^{\rho}f_{\rho\nu)}-g_{(\mu}f_{\nu)\rho}h^{\rho}]-\xi^2[g_{\mu}g_{\nu}(h^2-\nabla_{\rho}h^{\rho}) +g_{(\mu}f_{\nu)\rho}h^{\rho}]+\\
    +\frac14(\la^{-1}-\xi^2)^2g_{\mu}g_{\nu}f^2+h_{\mu}h_{\nu}+\nabla_{\mu}h_{\nu}.
\end{multline}
Consequently, the scalar curvature of the metric $G_{\mu\nu}$ is written as
\begin{equation}
    \tilde{R}=G^{\mu\nu}\tilde{R}_{\mu\nu}=R+2\nabla_{\rho}h^{\rho}-\frac{1}{4}(\la^{-1}-\xi^2)f^2.
\end{equation}

\section{Relevant parts of the heat kernel expansion coefficients}\label{HKE_coeff}

Here we provide the parts of the heat kernel expansion coefficients relevant to the high-temperature expansion of the one-loop free energy in a three dimensional space. It is these coefficients that determine the divergent and finite parts of the high-temperature expansion when $\be\rightarrow0$. We just employ the known expressions \cite{VasilHeatKer,Avramid,Ven,Gilkey} for the heat kernel coefficients adapted to the operator $H_G$ \eqref{H_G}. Notice that we did not check the general formulas for the coefficients $a_k$ presented in \cite{VasilHeatKer,Avramid,Ven,Gilkey} save the terms at small $k$.

As follows from Eqs. \eqref{omega_b} and \eqref{omega_f}, the relevant parts of the heat kernel coefficients $a_k$ satisfy the inequality $j\geq k-4$ (for $d=3$), where $j$ is the power of $\omega$ entering the expression. In our case, using the notation of \cite{VasilHeatKer}, we should put
\begin{equation}
\begin{gathered}
    E\rightarrow V-E,\qquad\Omega_{\mu\nu}\rightarrow-i\omega f_{\mu\nu},\\
    V=\frac14 h^2-\frac12\nabla^\mu h_\mu,\qquad E=\frac{\omega^2}{\xi^2}-m^2.
\end{gathered}
\end{equation}
So, $E$ has the second power in $\omega$, the curvature of the gauge fields $\Omega$ is of the first power, and the other structures are of the zeroth order in $\omega$. Besides, as we resummed the heat kernel expansion (see Eq. \eqref{heat_kern_res}) and collected the terms $(V-E)$ and $\tilde R/6$ without derivatives to the exponent, such terms in $a_k$ must be obliterated. We should to warn the reader that the Riemann tensor used in \cite{VasilHeatKer} differs from our one by the sign, while the Ricci tensors coincide.

Denoting the relevant part of the coefficient by tilde, we have
\begin{equation}
    \tilde{a}_0=1,\qquad\tilde{a}_1=0,
\end{equation}
\begin{equation}\label{a_2}
\begin{array}{l}
    \tilde{a}_2=\frac{1}{2!}\Big[
    \frac{1}{15}\tilde{\nabla}^2\tilde{R}
    -\frac{1}{90}\tilde{R}_{\mu\nu}\tilde{R}^{\mu\nu}
    +\frac{1}{90}\tilde{R}_{\alpha\beta\mu\nu}\tilde{R}^{\alpha\beta\mu\nu}
    +\frac{1}{3}\tilde{\nabla}^2V
    -\frac{1}{3}\tilde{\nabla}^2E
    +\frac{1}{6}\Omega^2\Big],
\end{array}
\end{equation}
where
\begin{equation}
\begin{array}{l}
    \tilde{\nabla}^2\tilde{R}=(\nabla^2-h^\rho\nabla_\rho)[R+2\nabla^\mu h_\mu-\frac14(\la^{-1}-\xi^2)f^2],\\
    \tilde{R}^{\mu\nu}\tilde{R}_{\mu\nu}=R^{\mu\nu}R_{\mu\nu}+2R^{\mu\nu}[h_\mu h_\nu+\nabla_\mu h_\nu+\frac12(\la^{-1}-\xi^2)f^2_{\mu\nu}]+\frac14(\la^{-1}-\xi^2)^2[\Tr f^4+\frac14(f^2)^2]+\\
    +\frac12(\la^{-1}-\xi^2)[(\nabla^\mu f_{\mu\nu})^2+hf^2h+2\nabla^\mu h^\nu f^2_{\mu\nu}]
    +(\la^{-1}+\xi^2)\nabla^\rho f_{\rho\mu}f^{\mu\nu}h_\nu+\\
    +\frac{\xi^2}{2}f^2(\nabla^\mu h_\mu -h^2)
    +h^\mu\nabla_\mu h^2+2h^2\nabla^\mu h_\mu+\nabla_\mu h_\nu \nabla^\mu h^\nu-(\nabla^\mu h_\mu)^2,\\
    \tilde R^{\mu\nu\rho\s}\tilde R_{\mu\nu\rho\s}=R^{\mu\nu\rho\s}R_{\mu\nu\rho\s} -\frac32(\la^{-1}-\xi^2)R^{\rho\nu\s\mu}f_{\rho\nu}f_{\s\mu}+(\la^{-1}-\xi^2)(\nabla_\rho f_{\mu\s}\nabla^\rho f^{\mu\s}+2hf^2h)-\\
    -2\xi^2(\nabla^\mu h^\nu f^2_{\mu\nu}-hf^2h+\frac32h^\mu\nabla_\mu f^2+3h^2f^2+2g^2\nabla_\mu h_\nu \nabla^\mu h^\nu+2g^2h^\mu\nabla_\mu h^2)+\\
    +\frac18(\la^{-1}-\xi^2)^2[3(f^2)^2+5\Tr f^4],\\
    \tilde{\nabla}^2V=(\nabla^2-h^\rho\nabla_\rho)(\frac14 h^2-\frac12\nabla^\mu h_\mu),\\
    \tilde{\nabla}^2E=\omega^2g^2(6h^2-2\nabla^{\mu}h_{\mu}),\\
    \Omega^2\equiv\Omega^{\mu\nu}\Omega_{\mu\nu}=-\omega^2f^2.
\end{array}
\end{equation}
In the expression \eqref{a_2} for $\tilde{a}_2$ and further for $\tilde{a}_k$ (Eqs. \eqref{a_3}, \eqref{a_4}, \eqref{a_5}, and \eqref{a_6}), the indices are raised and lowered by the use of the metric $G_{\mu\nu}$. Collecting all the above terms, we obtain
\begin{equation}
\begin{array}{l}
    \tilde{a}_2=
    \frac{1}{180}\Big[R^{\mu\nu\rho\s}R_{\mu\nu\rho\s}+\frac{3\xi^2}2R^{\rho\nu\s\mu}f_{\rho\nu}f_{\s\mu} -R^{\mu\nu}(R_{\mu\nu}+2h_\mu h_\nu+2\nabla_\mu h_\nu-\xi^2 f^2_{\mu\nu}) +6(\nabla^2-h^\mu\nabla_\mu)R\\
    +\frac{\xi^4}{16}(5(f^2)^2+6\Tr f^4)+\frac{\xi^2}{2}(\nabla^\mu f_{\mu\nu})^2-\xi^2\nabla_\mu f_{\nu\rho}\nabla^\mu f^{\nu\rho} -\xi^2\nabla^\rho f_{\rho\mu}f^{\mu\nu}h_\nu+\frac{\xi^2}2hf^2h -\xi^2\nabla^\mu h^\nu f^2_{\mu\nu}\\
    +\frac{5\xi^2}2 f^2(\nabla^\mu h_\mu -h^2) -2h^2\nabla^\mu h_\mu +(\nabla^\mu h_\mu)^2 +10\nabla_\mu h_\nu\nabla^\mu h^\nu-\frac{25}{2}h^\mu\nabla_\mu h^2+15 h^\mu\nabla^2 h_\mu -3\nabla^2\nabla^\mu h_\mu\\
    +3h^\mu\nabla_\mu\nabla_\nu h^\nu+\frac{3\xi^2}2 (h^\mu\nabla_\mu f^2+\nabla^2 f^2)
    \Big]+\omega^2 g^2\Big[\frac13\nabla^\mu h_\mu-h^2-\frac{\xi^2}{12}f^2\Big].
\end{array}
\end{equation}
The relevant part of the next heat kernel expansion coefficient is (see also \cite{Avramid,Gilkey})
\begin{equation}\label{a_3}
\begin{array}{l}
    \tilde{a}_3=\frac{1}{3!}\Big[
    \frac{2}{15}\tilde{\nabla}_{\mu}\Omega_{\nu\rho}\tilde{\nabla}^{\mu}\Omega^{\nu\rho}
    +\frac{1}{30}(\tilde{\nabla}^{\nu}\Omega_{\nu\mu})^2
    +\frac{1}{5}\Omega^{\mu\nu}\tilde{\nabla}^2\Omega_{\mu\nu}
    +\frac{1}{10}\tilde{R}_{\alpha\beta\mu\nu}\Omega^{\alpha\beta}\Omega^{\mu\nu}
    +\frac{1}{15}\tilde{R}^{\mu\nu}\Omega^2_{\mu\nu}-\\
    -\frac{1}{10}(\tilde{\nabla}^2)^2E-
    \tilde{\nabla}_{\mu}V\tilde{\nabla}^{\mu}E
    -\frac{1}{15}\tilde{R}_{\mu\nu}\tilde{\nabla}^{\mu}\tilde{\nabla}^{\nu}E
    -\frac{1}{5}\tilde{\nabla}^{\mu}\tilde{R}\tilde{\nabla}_{\mu}E
    +\frac{1}{2}\tilde{\nabla}_{\mu}E\tilde{\nabla}^{\mu}E \Big],
\end{array}
\end{equation}
where
\begin{equation}
\begin{array}{l}
    \tilde{\nabla}_{\mu}\Omega_{\nu\rho}\tilde{\nabla}^{\mu}\Omega^{\nu\rho}=
    -\omega^2[\nabla_{\mu}f_{\nu\rho}\nabla^{\mu}f^{\nu\rho}+2hf^2h+\frac12(\la^{-1}-\xi^2)\Tr f^4],\\
    (\tilde{\nabla}^{\nu}\Omega_{\nu\mu})^2=-\omega^2[(\nabla^{\nu}f_{\nu\mu})^2+\frac{1}{4}(\lambda^{-1}-\xi^2)(f^2)^2+2\nabla^\nu f_{\nu\mu}f^{\mu\rho}h_\rho-hf^2h],\\
    \Omega^{\mu\nu}\tilde{\nabla}^2\Omega_{\mu\nu}=-\omega^2[f^{\mu\nu}\nabla^2f_{\mu\nu}-\frac12(\la^{-1}-\xi^2)\Tr f^4-2hf^2h-\frac12h^\rho\nabla_\rho f^2],\\
    \tilde{R}_{\alpha\beta\mu\nu}\Omega^{\alpha\beta}\Omega^{\mu\nu}=-\omega^2[R_{\alpha\beta\mu\nu}f^{\alpha\beta}f^{\mu\nu}-\frac{1}{2}(\lambda^{-1}-\xi^2)((f^2)^2+\Tr f^4)],\\
    (\tilde{\nabla}^2)^2E=-2\omega^2g^2[\nabla^2\nabla^\mu h_\mu-6\nabla_\mu h_\nu\nabla^\mu h^\nu-6h^\mu\nabla^2 h_\mu+15h^\mu\nabla_\mu h^2-5h^\mu\nabla_\mu\nabla_\nu h^\nu-2(\nabla^\mu h_\mu-3h^2)^2],\\
    \tilde{R}^{\mu\nu}\Omega^2_{\mu\nu}=-\omega^2[R^{\mu\nu}f^2_{\mu\nu}+
    \frac{1}{2}(\lambda^{-1}-\xi^2)\Tr f^4+hf^2h+f^2_{\mu\nu}\nabla^{\mu}h^{\nu}],\\
    \tilde{\nabla}_{\mu}V\tilde{\nabla}^{\mu}E=\omega^2g^2[h^{\mu}\nabla_{\mu}\nabla_{\nu}h^{\nu}-
    \frac12h^{\mu}\nabla_{\mu}h^2],\\
    \tilde{R}_{\mu\nu}\tilde{\nabla}^{\mu}\tilde{\nabla}^{\nu}E=\omega^2g^2[R_{\mu\nu}(4h^\mu h^\nu-2\nabla^\mu h^\nu)+(3\la^{-1}-\xi^2)hf^2h+6h^4+h^\mu\nabla_\mu h^2-2\nabla_\mu h_\nu\nabla^\mu h^\nu-\\
    -2h^2\nabla^\mu h_\mu-(\la^{-1}-\xi^2)(f^2_{\mu\nu}\nabla^\mu h^\nu+\nabla^\mu f_{\mu\nu}f^{\nu\rho}h_\rho)+\frac12h^2\xi^2f^2],\\
    \tilde{\nabla}^{\mu}\tilde{R}\tilde{\nabla}_{\mu}E=-2\omega^2g^2h^\nu\nabla_\nu[R+2\nabla^\mu h_\mu-\frac14(\la^{-1}-\xi^2)f^2],\\
    \tilde{\nabla}_{\mu}E\tilde{\nabla}^{\mu}E=4\omega^4g^4h^2.
\end{array}
\end{equation}
And so
\begin{equation}
\begin{array}{l}
    \tilde{a}_3=-\frac{\omega^2 g^2}{90}\Big[
    \frac{3\xi^2}{2}R^{\al\be\mu\nu}f_{\al\be}f_{\mu\nu} +R^{\mu\nu}(\xi^2f^2_{\mu\nu}+4h_\mu h_\nu-2\nabla_\mu h_\nu) +\frac{3\xi^4}{4}(\Tr f^4+\frac56(f^2)^2)\\
    +2\xi^2\nabla_\mu f_{\nu\rho}\nabla^\mu f^{\nu\rho}
    +\frac{\xi^2}{2}(\nabla^\nu f_{\nu\mu})^2 +3\xi^2 f^{\mu\nu}\nabla^2f_{\mu\nu} -\frac{5\xi^2}2 hf^2h +2\xi^2\nabla^\nu f_{\nu\mu}f^{\mu\rho}h_\rho+3\xi^2 h^\mu\nabla_\mu f^2\\
    +2\xi^2\nabla^\mu h^\nu f^2_{\mu\nu} -3\nabla^2\nabla^\mu h_\mu
    +16\nabla^\mu h^\nu\nabla_\mu h_\nu +18 h^\mu\nabla^2 h_\mu -\frac{103}2h^\mu\nabla_\mu h^2 +18h^\mu\nabla_\mu\nabla_\nu h^\nu +6(\nabla^\mu h_\mu)^2\\
    -38h^2\nabla^\mu h_\mu
    +60 h^4
    -\frac{5\xi^2}{2}h^2f^2 -6h^\mu\nabla_\mu R\Big]
    +\frac13\omega^4 g^4 h^2.
\end{array}
\end{equation}
If we had included the nonminimal coupling $\kappa R$ of the scalar field with gravity to the Klein-Gordon equation \eqref{KG_eq}, this just would have resulted in a change of the coefficients of the scalar curvature entering the expressions for $\tilde{a}_2$ and $\tilde{a}_3$ and a redefinition of the effective mass $\tilde{m}$:
\begin{equation}\label{nonmin}
    \tilde{m}^2\rightarrow\tilde{m}^2-\kappa R,\qquad\tilde{a}_2\rightarrow\tilde{a}_2-\frac{\kappa}{6}(\nabla^2-h^\mu\nabla_\mu)R,\qquad\tilde{a}_3\rightarrow\tilde{a}_3-\frac{\omega^2g^2\kappa}{3}h^\mu\nabla_\mu R.
\end{equation}

The higher coefficients are encrypted in \cite{Ven}. Notice that formula (6.6) for $\Zs_{(2)}$ in \cite{Ven} contains a spurious term $R^2/12$. Making use of the notation from that paper, we write out the relevant parts of the coefficients:
\begin{equation}
\begin{array}{l}
    \tilde{a}_4=\frac{1}{4!}\Big[
    \frac25 \Zs_{(1}^\dag\hat{\Zs}\Zs_{1)}
    +\frac25\{\Zs^\dag_{(1}\Zs_{3)}\}
    +\frac25\Zs_{(2}^\dag\Zs_{2)}
    +\frac15\Zs^2_{\br2}
    \Big],
\end{array}
\end{equation}
where
\begin{equation}
\begin{array}{l}
    \Zs_{(1}^\dag\hat{\Zs}\Zs_{1)}\approx-\frac13\tilde{R}^{\mu\nu}X_\mu X_\nu+\frac23X^\mu\Omega_{\mu\nu}\Omega^{\nu\la}_{\ \ ;\la},\\
    \{\Zs^\dag_{(1}\Zs_{3)}\}\approx\frac43 X^\mu X^{\ \nu}_{\mu\ \nu}+\frac23 X^\mu X^\nu_{\ \nu\mu}+X^\mu(\Omega^2)_{;\mu}-\frac43X^\mu\Omega_{\mu\nu}\Omega^{\nu\la}_{\ \ ;\la},\\
    \Zs_{(2}^\dag\Zs_{2)}\approx\frac13[(X^\mu_{\ \mu})^2+2X^{\mu\nu}X_{\mu\nu}-2X^{\mu\nu}\Omega^2_{\mu\nu}+X^\mu_{\ \mu}\Omega^2+\frac14(\Omega^2)^2+\frac12\Tr\Omega^4],\\
    \Zs^2_{\br2}\approx(X^\mu_{\ \mu})^2+X^\mu_{\ \mu}\Omega^2+\frac14(\Omega^2)^2.
\end{array}
\end{equation}
Here $X=V-E$ as above, the indices of $X$ and semicolon denote the covariant differentiation with respect to the Levi-Civita connection $\tilde{\Ga}^\rho_{\mu\nu}$, and the approximate equality means that the relevant part of the expression is retained. Therefore we have (see also \cite{Avramid})
\begin{equation}\label{a_4}
\begin{array}{l}
    \tilde{a}_4=\frac{1}{4!}\Big[
    \frac25\tilde{R}_{\mu\nu}\tilde{\nabla}^\mu E \tilde{\nabla}^\nu E
    +\frac4{15}\tilde{\nabla}^\mu\Omega_{\mu\nu}\Omega^{\nu\rho}\tilde{\nabla}_\rho E
    +\frac45\tilde{\nabla}^\mu E\tilde{\nabla}_\mu\tilde{\nabla}^2E
    -\frac25\tilde{\nabla}^\mu E\tilde{\nabla}_\mu\Omega^2
    +\frac13(\tilde{\nabla}^2E)^2-\\
    -\frac13\tilde{\nabla}^2E\Omega^2
    +\frac1{12}(\Omega^2)^2
    +\frac4{15}\tilde{\nabla}^\mu\tilde{\nabla}^\nu E\tilde{\nabla}_\mu\tilde{\nabla}_\nu E
    +\frac4{15}\tilde{\nabla}^\mu\tilde{\nabla}^\nu E\Omega^2_{\mu\nu}
    +\frac1{15}\Tr\Omega^4\Big],
\end{array}
\end{equation}
where
\begin{equation}
\begin{array}{l}
    \tilde{R}_{\mu\nu}\tilde{\nabla}^{\mu}E\tilde{\nabla}^{\nu}E=2\omega^4g^4[2R_{\mu\nu}h^{\mu}h^{\nu}+
    (\lambda^{-1}-\xi^2)hf^2h+2h^4+h^{\mu}\nabla_{\mu}h^2],\\
    \tilde{\nabla}^{\mu}\Omega_{\mu\nu}\Omega^{\nu\rho}\tilde{\nabla}_{\rho}E=2\omega^4g^2(\nabla^{\mu}f_{\mu\nu}f^{\nu\rho}h_\rho-hf^2h),\\
    \tilde{\nabla}^{\mu}E\tilde{\nabla}_{\mu}\tilde{\nabla}^2E=4\omega^4g^4
    (h^{\mu}\nabla_{\mu}\nabla_{\nu}h^{\nu}-2h^2\nabla^{\mu}h_{\mu}+6h^4-3h^{\mu}\nabla_{\mu}h^2),\\
    \tilde{\nabla}^{\mu}E\tilde{\nabla}_{\mu}\Omega^2=2\omega^4g^2h^{\mu}\nabla_{\mu}f^2,\\
    (\tilde{\nabla}^2E)^2=4\omega^4g^4(3h^2-\nabla_{\mu}h^{\mu})^2,\\
    \tilde{\nabla}^2E\Omega^2=2\omega^4g^2f^2(\nabla_{\mu}h^{\mu}-3h^2),\\
    (\Omega^2)^2=\omega^4(f^2)^2,\\
    \tilde{\nabla}^{\mu}\tilde{\nabla}^{\nu}E\tilde{\nabla}_{\mu}\tilde{\nabla}_{\nu}E=4\omega^4g^4[
    \nabla_\mu h_\nu\nabla^\mu h^\nu-2h^\mu\nabla_\mu h^2+3h^4-\frac12(\lambda^{-1}-\xi^2)hf^2h],\\
    \tilde{\nabla}^{\mu}\tilde{\nabla}^{\nu}E\Omega^2_{\mu\nu}=2\omega^4g^2(\nabla^{\mu}h^{\nu}f^2_{\mu\nu}-2hf^2h),\\
    \Tr\Omega^4=\omega^4\Tr f^4.
\end{array}
\end{equation}
Summing up, we obtain
\begin{equation}
\begin{array}{l}
    \tilde{a}_4=\frac{\omega^4 g^4}{15}\Big[
    R_{\mu\nu}h^\mu h^\nu +\frac{\xi^4}{96}(5(f^2)^2+4\Tr f^4) -\frac{7\xi^2}6hf^2h +\frac{\xi^2}{3}\nabla^\mu h^\nu f^2_{\mu\nu} +\frac{\xi^2}3\nabla^\mu f_{\mu\nu}f^{\nu\rho}h_\rho -\frac{\xi^2}2h^\mu\nabla_\mu f^2\\
    -\frac{5\xi^2}{12}f^2(\nabla^\mu h_\mu-3h^2) +\frac{45}2 h^4 -\frac{41}{6}h^\mu\nabla_\mu h^2 +2h^\mu\nabla_\mu\nabla_\nu h^\nu -9h^2\nabla^\mu h_\mu +\frac56(\nabla^\mu h_\mu)^2 +\frac23\nabla_\mu h_\nu\nabla^\mu h^\nu
    \Big].
\end{array}
\end{equation}
Further,
\begin{equation}\label{a_5_pre}
\begin{array}{l}
    \tilde{a}_5=\frac{1}{5!}\Big[
    \frac13\Zs^\dag_{(1}\hat{\Zs}{}^2\Zs_{1)}
    +\frac97\Zs^\dag_{(1}\hat{\Zs}_2\Zs_{1)}
    +\frac67\{\Zs^\dag_{(1}\hat{\Zs}_1\Zs_{2)}\}
    +\frac13\{\Zs^\dag_{(1}\Zs_{1)}\Zs_{\br2}\}
    \Big],
\end{array}
\end{equation}
where
\begin{equation}\label{Z_5}
\begin{array}{l}
    \Zs^\dag_{(1}\hat{\Zs}{}^2\Zs_{1)}\approx X^\mu\Omega^2_{\mu\nu}X^\nu,\\
    \Zs^\dag_{(1}\hat{\Zs}_2\Zs_{1)}\approx\frac23X^\mu X_{\mu\nu}X^\nu+\frac13X^\mu X_\mu X^\nu_{\ \nu}+\frac16X^\mu X_\mu\Omega^2-\frac13X^\mu\Omega^2_{\mu\nu}X^\nu,\\
    \{\Zs^\dag_{(1}\hat{\Zs}_1\Zs_{2)}\}\approx2\Zs^\dag_{(1}\hat{\Zs}_2\Zs_{1)},\\
    \{\Zs^\dag_{(1}\Zs_{1)}\Zs_{\br2}\}\approx2[X^\mu X_\mu X^\nu_{\ \nu}+\frac12X^\mu X_\mu\Omega^2].
\end{array}
\end{equation}
Hence, substituting \eqref{Z_5} to \eqref{a_5_pre}, we come to
\begin{equation}\label{a_5}
\begin{array}{l}
    \tilde{a}_5=\frac{1}{5!}\Big[
    -2\tilde{\nabla}^{\mu}E\tilde{\nabla}^{\nu}E\tilde{\nabla}_{\mu}\tilde{\nabla}_{\nu}E
    -\frac{5}{3}\tilde{\nabla}^{\mu}E\tilde{\nabla}_{\mu}E\tilde{\nabla}^2E
    +\frac{5}{6}\tilde{\nabla}^{\mu}E\tilde{\nabla}_{\mu}E\Omega^2
    -\frac23\tilde{\nabla}^{\mu}E\tilde{\nabla}^{\nu}E\Omega^2_{\mu\nu}
    \Big],
\end{array}
\end{equation}
where
\begin{equation}
\begin{array}{l}
    \tilde{\nabla}^{\mu}E\tilde{\nabla}^{\nu}E\tilde{\nabla}_{\mu}\tilde{\nabla}_{\nu}E=
    4\omega^6g^6(4h^4-h^{\mu}\nabla_{\mu}h^2),\\
    \tilde{\nabla}^{\mu}E\tilde{\nabla}_{\mu}E\tilde{\nabla}^2E=
    8\omega^6g^6h^2(3h^2-\nabla^{\mu}h_{\mu}),\\
    \tilde{\nabla}^{\mu}E\tilde{\nabla}_{\mu}E\Omega^2=-4\omega^6g^4h^2f^2,\\
    \tilde{\nabla}^{\mu}E\tilde{\nabla}^{\nu}E\Omega^2_{\mu\nu}=-4\omega^6g^4hf^2h.
\end{array}
\end{equation}
Collecting all the terms together, we get
\begin{equation}
\begin{array}{l}
    \tilde{a}_5=-\frac{4\omega^6g^6}{5!}\Big[
    18h^4
    -2h^\mu\nabla_\mu h^2
    -\frac{10}3 h^2\nabla^\mu h_\mu
    +\frac{5\xi^2}6h^2f^2
    -\frac{2\xi^2}3hf^2h
    \Big].
\end{array}
\end{equation}
The last coefficient that we need is
\begin{equation}\label{a_6}
\begin{array}{l}
    \tilde{a}_6=\frac{1}{6!}\Big[\frac47 X_{(1}X_{1)}X_{(1}X_{1)}+\frac{27}{14}X_{(1}X_{1}X_{1}X_{1)}\Big]\approx\frac{1}{6!}\frac52(\tilde{\nabla}^{\mu}E\tilde{\nabla}_{\mu}E)^2
    =\frac{40}{6!}\omega^8g^8h^4.
\end{array}
\end{equation}

In conclusion we give the explicit expression for the finite part of the high-temperature expansion of the one-loop contribution to the free energy (see Eqs. \eqref{omega_b_fin}, \eqref{omega_f_fin}) retaining only the terms at the nonegative powers of $m^2$:
\begin{equation}\label{fin_part}
\begin{array}{l}
   \tilde{m}^4-\tilde{m}^2(\frac23\tilde{a}_2^{\br2}+\frac53\tilde{a}_3^{\br4})-\frac43(\tilde{a}_3^{\br2}+2\tilde{a}_4^{\br4}+\frac{23}4\tilde{a}_5^{\br6}+22\tilde{a}_6^{\br8})=m^4 +m^2(\frac13 R +\frac{5\xi^2}{36}f^2 -\frac59 \nabla^\mu h_\mu +\frac{11}{18}h^2)\\
   +\frac1{45}\Big[\xi^2 R^{\al\be\mu\nu}f_{\al\be}f_{\mu\nu} +\frac23R^{\mu\nu}(\xi^2f^2_{\mu\nu}-2\nabla_\mu h_\nu- 2h_\mu h_\nu)+\frac{\xi^4}{6}(\Tr f^4 +\frac{35}{32}(f^2)^2) +\frac{\xi^2}{3}(\nabla^\mu f_{\mu\nu})^2\\
   +\frac{4\xi^2}{3}(\nabla_\mu f_{\nu\rho}\nabla^\mu f^{\nu\rho} -\nabla^\nu f_{\nu\mu}f^{\mu\rho}h_\rho -f^2_{\mu\nu}\nabla^\mu h^\nu) +2\xi^2(f^{\mu\nu}\nabla^2f_{\mu\nu} +h^\mu\nabla_\mu f^2) -\frac{5\xi^2}{16}h^2f^2
   +\frac{15\xi^2}{8}f^2\nabla^\mu h_\mu\\
   -2\nabla^2\nabla^\mu h_\mu +\frac{16}{3}\nabla^\mu h^\nu\nabla_\mu h_\nu +8h^\mu\nabla^2 h_\mu +\frac14(\nabla^\mu h_\mu)^2 +\frac54h^2\nabla^\mu h_\mu -\frac83h^\mu\nabla_\mu h^2 -\frac{109}{48}h^4\Big]\\
   -\frac{4}{45}h^\mu\nabla_\mu R +\frac{1}{6}R\Big(\frac{11}{18}h^2 -\frac{5}{9}\nabla^\mu h_\mu +\frac{5\xi^2}{36}f^2\Big) +\frac{1}{36}R^2.
\end{array}
\end{equation}
The nonminimal coupling adds to this part the following terms:
\begin{equation}
    -\kappa R(\frac{11}{18}h^2-\frac59\nabla^\mu h_\mu +\frac{5\xi^2}{36}f^2+2m^2)+(\kappa^2-\frac{\kappa}{3})R^2+\frac{4\kappa}{9}h^\mu\nabla_\mu R.
\end{equation}
In the conformal case, $m^2=0$, $\kappa=1/6$, this contribution cancels completely the terms proportional to the scalar curvature $R$ (without derivatives) entering the finite part.

%\newpage
\begin{acknowledgments}

We are grateful to D.~V. Vassilevich for the fruitful correspondence elucidating certain aspects of the heat kernel expansion technique. ISK appreciates the Dynasty Foundation for financial support. The work is done partially under the project 2.3684.2011 of Tomsk State University. It is also supported by the RFBR grants No. 12-02-31071-mol-a and No. 13-02-00551, and by the Russian Ministry of Education and Science, contracts No. 14.B37.21.0911 and No. 14.B37.21.1298.

\end{acknowledgments}

\end{document}